\documentclass[10pt,a4paper,english,aps,prb,longbibliography,reprint,twocolumn]{revtex4-1}
\usepackage[T1]{fontenc}
\usepackage{babel}
\usepackage{mathtools}
\usepackage{amsmath}
\usepackage{graphicx}
\usepackage[unicode=true,pdfusetitle,
 bookmarks=true,bookmarksnumbered=false,bookmarksopen=false,
 breaklinks=false,pdfborder={0 0 1},backref=false,colorlinks=false]
 {hyperref}

\makeatletter

\pdfpageheight\paperheight
\pdfpagewidth\paperwidth

\newcommand{\noun}[1]{\textsc{#1}}

\usepackage{subfigure}
\usepackage{graphicx}

\makeatother

\begin{document}

\title{Charge-Kondo effect mediated by repulsive interactions}

\author{S. Mojtaba Tabatabaei}

\affiliation{Department of Physics, Shahid Beheshti University, Tehran, Iran}
\begin{abstract}
We investigate the Kondo effect in a double-quantum-dot which is capacitively
coupled to a charge-Qubit. It is shown that due to this capacitive
coupling, the bare inter-dot repulsive interaction in the double-quantum-dot
is effectively reduced and eventually changed to an attractive interaction
for strong couplings between the double-quantum-dot and the Qubit.
By deriving the low-energy effective Hamiltonian of the system, we
find that the low energy dynamics of the system corresponding to these
two positive or negative effective interaction regimes can be described,
respectively, by an isotropic orbital-Kondo or an anisotropic charge-Kondo
Hamiltonian. Moreover, we study various thermodynamic and electronic
transport properties of the system by using the numerical renormalization
group method.
\end{abstract}

\date{\today}
\maketitle

\section{Introduction}

During last two decades, the Kondo effect in quantum dots(QD) has
become an attractive research field in condensed matter physics\cite{hewson1997kondo,odom2000magnetic,buitelaar2002quantum,yu2004kondo,potok2007observation,fu2007manipulating,calvo2009kondo,pillet2010andreev,lee2014spin,iftikhar2015two}.
The main motivation for these studies is the possibility to explore
various aspects of such an important effect in an experimentally accessible
and fully tunable manner. Theoretically, Kondo effect is expected
to be arisen at low temperatures whenever a localized system with
a degenerate ground state is coupled to an environment with the same
degeneracy and it is usually referred to the formation of a zero energy
resonance state in the density of states of the system as a result
of higher order tunneling processes between the localized system and
the environment\cite{hewson1997kondo}. Depending on the nature of
the degrees of freedom contributing to the manifestation of this effect,
different kinds of Kondo effects have been identified in the literature.
For example, the \emph{spin}-Kondo effect is the most addressed Kondo
effect which is associated with the fluctuations in the degenerate
spin-up and down states in the local system\cite{inoshita1998kondo,cronenwett1998tunable,Pustilnik2004}.
Another example is the \emph{orbital}-Kondo effect which is formed
in a double-quantum-dot(DQD) and associated with the degenerate pseudo-spin
states corresponding to the occupation of the DQD by an electron in
its left or right dot\cite{jarillo2005orbital,Amasha2013,Bao2014}.

The other somehow elusive Kondo effect is the \emph{charge}-Kondo
effect which is associated with the fluctuations in degenerate states
with different charge occupations. In this sense, the charge-Kondo
effect is expected tooccur in negative-$U$ centers where an attractive
interaction makes the doubly occupied or empty states have lower energies
than the singly occupied states. Despite the fact that the theory
of charge-Kondo effect was developed in early 1990s\cite{taraphder1991heavy},
its experimental realization has not been reached until recent years.
It was first reported in the bulk $PbTe$ semiconductors doped with
$Tl$ valence skipping elements which are in essence acting as negative-$U$
centers in the host material\cite{Matsushita2005,Dzero2005,matusiak2009evidence,Costi2012}.
Another observation of the charge-Kondo effect was also reported in
transport through single electron transistors formed at the $LaAlO_{3}/SrTiO_{3}$
interfaces\cite{cheng2015electron,prawiroatmodjo2017transport,fang2017charge}.
Meanwhile, some efforts have been also devoted to engineer the attractive
interaction required for the charge-Kondo effect by introducing other
degrees of freedom interacting with electrons. Refs.{[}\onlinecite{Borda2003,alexandrov2003bistable,Cornaglia2004,Galpin2005,Liu2010,garate2011charge,andergassen2011mechanism,Yoo2014,szechenyi2017electron}{]}
are examples of such theoretical proposals. 

Recently, Hamo et.al\cite{Hamo2016} reported the observation of attracting
electrons in a setup composed of a carbon nanotube double quantum
dot(DQD) along with a charge-Qubit both of which are constructed on
separate microchips, placed one above the other and perpendicular
to each other. They found that an attractive interaction will be induced
between the electrons in the DQD when the height of the Qubit above
the DQD becomes less than a specific value. The attraction mechanism
in their setup, as is discussed by Little\cite{little1964possibility},
is purely electronic and of excitonic origin: An electron on the DQD
repels the electron in the up state of the Qubit leaving behind it
a cloud of positive charge on the Qubit which in turn, this positive
charge will attract the other electron on the DQD making the whole
as two attracting electrons on the DQD.

In view of the presence of attractive electrons in the DQD-Qubit system,
it is natural to ask whether the electronic transport through DQD
could show the charge-Kondo effect or not? It should be emphasized
that in their experiment, Hamo et.al, observed a conductance enhancement
on the degeneracy points between the two empty and doubly occupied
states of the DQD, a signature which was assigned to the charge-Kondo
effect by them. However, so far, there has not been presented a rigorous
description for the Kondo effect in the DQD-Qubit coupled system in
the literature.

In the present work, we theoretically explore the characteristics
of the Kondo effect in the DQD-Qubit coupled system. In the rest of
the paper, in Sec.\ref{sec:The-Model}, we at first introduce the
model Hamiltonian of the DQD-Qubit system and discuss the effective
interaction induced in the DQD due to its coupling with the Qubit.
Then, in Sec.\ref{sec:Low-energy-Effective-Hamiltonian}, we derive
the low-energy effective Hamiltonian of the system by using the Raleigh-Schr\"{o}dinger
degenerate perturbation theory\cite{lim2010transport,garate2011charge}.
We show that for positive effective interaction in the DQD, the system
exhibits isotropic orbital-Kondo effect while in the negative effective
interaction regime, the anisotropic charge-Kondo effect will be manifested
in the system at low enough temperatures. Then in Sec.\ref{sec:Numerical-renormalization-group},
we supplement our study with the results obtained by using the numerical
renoramlization group method(NRG)\cite{bulla2008numerical} to confirm
the presence of these different Kondo effects in the system and extract
some of the DQD's transport properties in the charge-Kondo regime.
Finally, in Sec.\ref{sec:Conclusions}, we give the conclusions.

\section{\label{sec:The-Model}The Model}

\begin{figure}
\includegraphics[width=8.6cm]{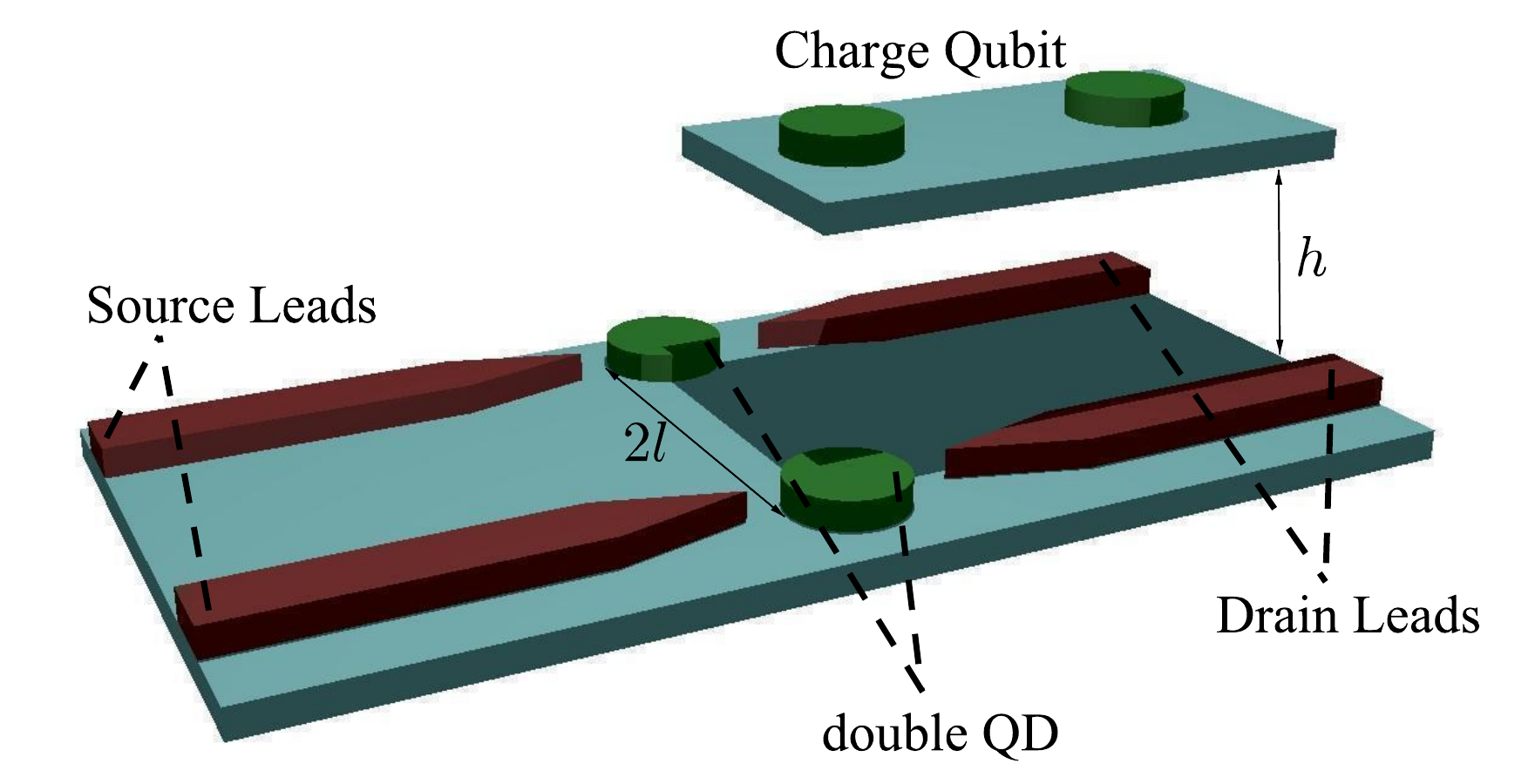}

\caption{\label{figsys}The charge Qubit is suspended above the DQD at height
$h$ and midway between the two dots of the DQD. Each dot of the DQD
is coupled to its own source and drain electrodes and the distance
between them is $2l$. One of the Qubit's dots has smaller distance
to the DQD than the second Qubit's dot and therefore, the capacitive
effect of the second dot of the Qubit on the dynamics of the DQD is
overestimated.}
\end{figure}
Our model system which is shown in Fig.\ref{figsys}, consists of
a parallel DQD and a charge Qubit. The Qubit is placed on a separate
host than DQD and is positioned above the mid-point between DQD at
height $h$, so that the DQD and the Qubit can interact with each
other only in a repulsive manner. The DQD is constituted from two
dots while each of them is contacted to its own source and drain electrodes.
The total Hamiltonian of the system is given by 
\begin{equation}
\hat{\mathcal{H}}=\hat{\mathcal{H}}_{S}+\hat{\mathcal{H}}_{L}+\hat{\mathcal{H}}_{T}.\label{eq:tot_Ham}
\end{equation}

The first term is the Hamiltonian of the DQD-Qubit system, which is
given by $\hat{\mathcal{H}}_{S}=\hat{\mathcal{H}}_{DQD}+\hat{\mathcal{H}}_{qubit}+\hat{\mathcal{H}}_{I}$.
The Hamiltonian of the DQD is given by 
\begin{equation}
\hat{\mathcal{H}}_{DQD}=\varepsilon_{1}\hat{n}_{1}+\varepsilon_{2}\hat{n}_{2}+U\hat{n}_{1}\hat{n}_{2},
\end{equation}
where $\hat{n}_{i}=\hat{d}_{i}^{\dagger}\hat{d}_{i}$, $\hat{d}_{i}^{\dagger}$
for $i=1,2$ creates an electron in the respective QD, $\varepsilon_{i}$
is the applied gate voltage and $U$ is the inter-dot electron-electron
interaction energy which is assumed to be a positive constant. In
order to capture the main physics of the charge-Kondo effect, we consider
the electrons to be spinless (by applying a large magnetic field)
and therefore, each QD can be occupied only by a single electron.
Moreover, $\hat{\mathcal{H}}_{qubit}$ denotes the Hamiltonian of
the Qubit which is given by 
\begin{equation}
\hat{\mathcal{H}}_{qubit}=-\frac{\omega_{0}}{2}\hat{\tau}_{z}+\frac{\Delta}{2}\hat{\tau}_{x},
\end{equation}
where $\omega_{0}$ is the energy difference between the energy levels
on either dots of the Qubit, $\Delta/2$ gives the corresponding electron
hybridization in the Qubit and $\hat{\tau}_{x}$ and $\hat{\tau}_{z}$
are the usual Pauli operators operating in the Qubit's Hilbert space
and defined respectively by $\hat{\tau}_{x}=(\left|\Uparrow\right\rangle \left\langle \Downarrow\right|+\left|\Downarrow\right\rangle \left\langle \Uparrow\right|)$
and $\hat{\tau}_{z}=(\left|\Uparrow\right\rangle \left\langle \Uparrow\right|-\left|\Downarrow\right\rangle \left\langle \Downarrow\right|)$,
where $\left|\Uparrow\right\rangle $ and $\left|\Downarrow\right\rangle $
represent the two charge sates of the Qubit. Furthermore, the interaction
Hamiltonian of the DQD-Qubit is given by 
\begin{equation}
\hat{\mathcal{H}}_{I}=\frac{\lambda}{2}\hat{n}_{d}\hat{\tau}_{z},
\end{equation}
where $\hat{n}_{d}=\hat{n}_{1}+\hat{n}_{2}$ and $\lambda$ is the
capacitive coupling energy between DQD and Qubit which is assumed
to be a positive constant. The value of $\lambda$ is proportional
to $r_{\lambda}^{-1}$, where $r_{\lambda}=\sqrt{l^{2}+h^{2}}$ is
the distance between the charge Qubit and the DQD (see Fig.\ref{figsys}
for the definitions of $l$ and $h$). 

The DQD is tunnel coupled to four normal metal electrodes so that
each dot coupled to its own source and drain electrodes. The electrodes
are described by $\hat{\mathcal{H}}_{L}=\sum_{i,j}\hat{\mathcal{H}}_{L,i,j}$,
where $j=S,D$ denote source and drain electrodes for the respective
dots $i=1,2$, and $\hat{\mathcal{H}}_{L,i,j}=\sum_{k}\varepsilon_{k}\,\hat{c}_{k,i,j}^{\dagger}\hat{c}_{k,i,j}$,
where $\hat{c}_{k,i,j}^{\dagger}$ is the corresponding operator for
electron creation with energy $\varepsilon_{k}$ in the respective
electrodes. The hybridization of each dot of the DQD with their electrodes
is assumed to be energy independent and characterized by a hybridization
constant $t$ and is described by $\hat{\mathcal{H}}_{T}=\sum_{k,i,j}t\,(\hat{c}_{k,i,,j}^{\dagger}\hat{d}_{i}+h.c)$.
The Hamiltonian of each source and drain electrode corresponding to
each dot may be transformed to the Hamiltonian of a single lead by
using the canonical transformations $\hat{\tilde{c}}_{k,i}=(\hat{c}_{k,i,S}+\hat{c}_{k,i,D})/\sqrt{2}$
and $\hat{\tilde{b}}_{k,i}=(\hat{c}_{k,i,S}-\hat{c}_{k,i,D})/\sqrt{2}$
and then, the expressions of the leads and the DQD-leads Hamiltonians
are given, respectively, by 
\begin{equation}
\hat{\mathcal{H}}_{L}=\sum_{k,i}\varepsilon_{k}\,\hat{\tilde{c}}_{k,i}^{\dagger}\hat{\tilde{c}}_{k,i},
\end{equation}
 and 
\begin{equation}
\hat{\mathcal{H}}_{T}=\sum_{k,i}t\,(\hat{\tilde{c}}_{k,i}^{\dagger}\hat{d}_{i}+h.c).
\end{equation}

In order to learn more about the impact of the capacitive interaction
between DQD and Qubit on the behavior of the system, it is useful
to calculate the eigenstates and their energies of the isolated DQD-Qubit
system. The eigenenergies of $\mathcal{H}_{S}$ are obtained straightforwardly
from $\mathcal{H}_{S}\left|\psi_{\pm}^{n_{1},n_{2}}\right\rangle =E_{\pm}^{n_{1},n_{2}}\left|\psi_{\pm}^{n_{1},n_{2}}\right\rangle $,
as 
\begin{equation}
E_{\pm}^{n_{1},n_{2}}=n_{1}\varepsilon_{1}+n_{2}\varepsilon_{2}+n_{1}n_{2}U\pm\frac{\Omega_{n_{d}}}{2},\label{eq:eigenenergy}
\end{equation}
where $n_{d}=n_{1}+n_{2}$ is the total occupation of DQD and $\Omega_{n}=\sqrt{\Delta^{2}+\left(\omega_{0}-n\lambda\right)^{2}}$.
Moreover, the eigenstates are found to be $\left|\psi_{\pm}^{n_{1},n_{2}}\right\rangle =\left|n_{1},n_{2}\right\rangle _{DQD}\left|n_{d},\pm\right\rangle _{qubit}$,
where $\left|n_{1},n_{2}\right\rangle _{DQD}$ is the occupation state
of DQD and $\left|n_{d},\pm\right\rangle _{qubit}$ is the Qubit eigenstate
given by 
\begin{equation}
\left|n_{d},\pm\right\rangle _{qubit}=b_{\pm}^{n_{d}}\left(\left(-\omega_{0}+n_{d}\lambda\pm\Omega_{n_{d}}\right)\left|\Uparrow\right\rangle +\Delta\left|\Downarrow\right\rangle \right),
\end{equation}
where $b_{\pm}^{n_{d}}$ is a normalization constant. Looking at the
eigenenergies, Eq.\eqref{eq:eigenenergy}, we can understand that
as a direct consequence of the coupling with Qubit, the electron-electron
interaction in DQD is reduced to an effective interaction equal to
\begin{align}
U_{eff} & =E_{-}^{1,1}+E_{-}^{0,0}-E_{-}^{1,0}-E_{-}^{0,1}\nonumber \\
 & =U+\Omega_{1}-\frac{1}{2}\left(\Omega_{2}+\Omega_{0}\right).\label{eq:Ueff}
\end{align}
More interestingly, we see that when $U<\frac{1}{2}\left(\Omega_{2}+\Omega_{0}\right)-\Omega_{1}$,
the sign of $U_{eff}$ becomes negative which means that in such situations,
there presents a net attractive interaction between the electrons
in the DQD. Even though it may seem surprising at first glance, we
can get some feelings of this attractive interaction by noting that
this is indeed an induced attraction between the electrons. In other
words, the presence of the oscillating polarization field of the Qubit
on the electrons in the DQD, dresses their electric potential and
forces them to favor the doubly occupied states more that the singly
ones which in turn can be considered as such that the two electrons
attract each other\cite{little1964possibility,Hamo2016}.

Another feature in Eq.\eqref{eq:eigenenergy} is that for some particular
parameter configurations, the ground-state of the system becomes degenerate,
composed of the two states $\bigl|\psi_{-}^{0,1}\bigr\rangle$ and
$\bigl|\psi_{-}^{1,0}\bigr\rangle$ ($\bigl|\psi_{-}^{0,0}\bigr\rangle$
and $\bigl|\psi_{-}^{1,1}\bigr\rangle$) in $U_{eff}>0$ ($U_{eff}<0$)
regime. Accordingly, we can expect that when the subsystem of DQD-Qubit
with the degenerate ground-state is appropriately coupled to the electrodes,
higher order electron tunneling processes between DQD and electrodes
dress this degenerate state and form a many-body Kondo resonance and
hence a Kondo effect arises in the system.

\section{Low-energy Effective Hamiltonian\label{sec:Low-energy-Effective-Hamiltonian}}

In order to identify the characteristics of the Kondo effect in our
model system, it is sufficient to derive a low-energy effective Hamiltonian
describing the low-temperature dynamics of the system up to the second-order
of the electron tunneling processes in the DQD. Here, we present the
main results and relegate the technical details to the Appendix \ref{sec:Derivation-of-effective}.
By using the projection operators 
\begin{equation}
\hat{P}_{\pm}^{n_{1},n_{2}}=\left|\psi_{\pm}^{n_{1},n_{2}}\right\rangle \left\langle \psi_{\pm}^{n_{1},n_{2}}\right|,
\end{equation}
we can calculate the effective Hamiltonian of the system by using\cite{lim2010transport,garate2011charge}
\begin{equation}
\hat{\mathcal{H}}_{eff}=\sum_{\mathclap{\substack{n_{1},n_{2}=0,1\\
\nu=\pm
}
}}\frac{\hat{P}_{0}\hat{\mathcal{H}}_{T}\hat{P}_{\nu}^{n_{1},n_{2}}\hat{\mathcal{H}}_{T}\hat{P}_{0}}{E_{0}-E_{\nu}^{n_{1},n_{2}}},\label{eq:heff}
\end{equation}
where $\hat{P}_{0}$ and $E_{0}$ are the corresponding projector
operator and energy of the ground state of the unperturbed system,
respectively. 

In positive $U_{eff}$ regime, the ground-state of the system can
become degenerate (which is essential for the Kondo effect to be arisen
in the system) and constituted from the two states $\bigl|\psi_{-}^{0,1}\bigr\rangle$
and $\bigl|\psi_{-}^{1,0}\bigr\rangle$ when the conditions $\lambda=\omega_{0}$,
$\varepsilon_{1}+\frac{U}{2}=\varepsilon_{2}+\frac{U}{2}=V_{g}$ and
$|V_{g}|<U_{eff}/2$, are satisfied. Then, using Eq.\eqref{eq:heff},
the low-energy effective Hamiltonian can be obtained by 
\begin{align}
\hat{\mathcal{H}}_{eff} & =\sum_{\mathclap{\substack{n_{1},n_{2}=0,1\\
\nu=\pm
}
}}\frac{\left(\hat{P}_{-}^{1,0}+\hat{P}_{-}^{0,1}\right)\hat{\mathcal{H}}_{T}\hat{P}_{\nu}^{n_{1},n_{2}}\hat{\mathcal{H}}_{T}\left(\hat{P}_{-}^{1,0}+\hat{P}_{-}^{0,1}\right)}{E_{-}^{1,0}-E_{\nu}^{n_{1},n_{2}}}\nonumber \\
 & \approx J\overrightarrow{S}_{d}.\overrightarrow{S}_{c},
\end{align}
where $\overrightarrow{S}_{d}$ and $\overrightarrow{S}_{c}$ are
the corresponding pseudo-spin vector of DQD and electrodes, respectively,
which are defined in Eq.\eqref{Sdef} and $J$ is the Kondo exchange
coupling constant which is given by
\begin{align}
J & =\frac{t^{2}}{\left(V_{g}+U\right)-\frac{\left(\lambda/2\right)^{2}}{\Delta+V_{g}+U}}-\frac{t^{2}}{V_{g}-\frac{\left(\lambda/2\right)^{2}}{V_{g}-\Delta}}.\label{eq:JI}
\end{align}
Thus, we see that the low-temperature dynamics of the system in $U_{eff}>0$
regime is governed by a $SU(2)$ isotropic orbital-Kondo Hamiltonian
in which a single electron on DQD plays the role of a pseudo-spin
making a singlet state with the corresponding pseudo spins of the
electrodes. The Kondo temperature which is the particular temperature
below which the Kondo effect is manifested in the system, is given
in this case by
\begin{equation}
T_{K}^{i}=\alpha\exp[-\frac{1}{\rho_{0}J}],\label{eq:tkI}
\end{equation}
where $\alpha$ is a proportionality constant and $\rho_{0}=1/\left(2D\right)$
is the density of states of the leads which is assumed to be constant
in the wide-band approximation and $D$ is the half band width of
the electrodes. To the lowest order in $\lambda$, the correction
to the value of $J$ equals to 
\begin{align}
J\approx-\frac{t^{2}U}{V_{g}\left(V_{g}+U\right)}+ & \frac{t^{2}\left(\lambda/2\right)^{2}}{\left(V_{g}+U\right){}^{2}\left(\Delta+V_{g}+U\right)}+\nonumber \\
 & \frac{t^{2}\left(\lambda/2\right)^{2}}{V_{g}^{2}\left(\Delta-V_{g}\right)}+\mathcal{O}(\lambda^{4}).
\end{align}
Hence, in the positive $U_{eff}$ regime, the DQD-Qubit coupling results
in an enhancement of the Kondo temperature of the system. Note that
for $\lambda=0$, Eq.\eqref{eq:JI} correctly reproduces the results
of the conventional orbital-Kondo effect in a DQD.

For negative $U_{eff}$ regime, by demanding the ground-state to be
degenerate and constituted from the two states $\bigl|\psi_{-}^{0,0}\bigr\rangle$
and $\bigl|\psi_{-}^{1,1}\bigr\rangle$, it can be found that the
the sufficient conditions are $\lambda=\omega_{0}$, $\varepsilon_{1}+\varepsilon_{2}+U=0$
and $|V_{z}|<-U_{eff}/2$ , where $V_{z}=\varepsilon_{1}+U/2=-(\varepsilon_{2}+U/2)$.
Now, we can perform a second-order perturbation to obtain the low-energy
effective Hamiltonian of the system as 
\begin{align}
\hat{\mathcal{H}}_{eff} & =\sum_{\mathclap{\substack{n_{1},n_{2}=0,1\\
\nu=\pm
}
}}\frac{\left(\hat{P}_{-}^{0,0}+\hat{P}_{-}^{1,1}\right)\hat{\mathcal{H}}_{T}\hat{P}_{\nu}^{n_{1},n_{2}}\hat{\mathcal{H}}_{T}\left(\hat{P}_{-}^{0,0}+\hat{P}_{-}^{1,1}\right)}{E_{-}^{0,0}-E_{\nu}^{n_{1},n_{2}}}\nonumber \\
 & \approx(J_{\parallel}I_{d}^{z}I_{c}^{z}+J_{\perp}(I_{d}^{x}I_{c}^{x}+I_{d}^{y}I_{c}^{y})),\label{eq:heffA}
\end{align}
where $\overrightarrow{I}_{d}(\overrightarrow{I}_{c})$ are the corresponding
iso-spin operators of the DQD(electrodes) which are defined by applying
a special particle-hole transformation, i.e $d_{2}^{\dagger}\rightarrow d_{2}$
and $\tilde{c}_{k,2}^{\dagger}\rightarrow-\tilde{c}_{k,2}$, on the
$\overrightarrow{S}_{d}(\overrightarrow{S}_{c})$ operators and are
given in Eq.\eqref{Idef}. Furthermore, the two parallel and transverse
exchange Kondo coupling constants $J_{\parallel}$ and $J_{\perp}$
are given by\begin{subequations}\label{erttt}
\begin{align}
J_{\parallel}= & 2t^{2}\Biggl[\frac{-\lambda+U+2V_{z}-\frac{\Delta^{2}\left(\lambda+2\Omega_{0}\right)}{\Delta^{2}+\lambda^{2}+\lambda\Omega_{0}}}{\left(U+2V_{z}-\Omega_{0}\right){}^{2}-\Delta^{2}}\nonumber \\
 & -\frac{\lambda-U+2V_{z}+\frac{\Delta^{2}\left(\lambda+2\Omega_{0}\right)}{\Delta^{2}+\lambda^{2}+\lambda\Omega_{0}}}{\left(-U+2V_{z}+\Omega_{0}\right){}^{2}-\Delta^{2}}\Biggr],\\
J_{\perp}= & \frac{2\Delta t^{2}}{\Omega_{0}}\Biggl(\frac{1}{2V_{z}-U+\frac{\lambda^{2}}{2V_{z}-U+2\Omega_{0}}}-\nonumber \\
 & \frac{1}{2V_{z}+U+\frac{\lambda^{2}}{2V_{z}+U-2\Omega_{0}}}\Biggr).
\end{align}
\end{subequations}Thus, the low-energy dynamics of the DQD-Qubit
system in the $U_{eff}<0$ regime is governed by an anisotropic charge-Kondo
Hamiltonian. In this case, the Kondo temperature is given by\cite{Yoo2014}
\begin{equation}
T_{K}^{a}=\beta\left(\frac{J_{\parallel}+\sqrt{J_{\parallel}^{2}-J_{\perp}^{2}}}{J_{\parallel}-\sqrt{J_{\parallel}^{2}-J_{\perp}^{2}}}\right)^{-\frac{1}{4\rho_{0}\sqrt{J_{\parallel}^{2}-J_{\perp}^{2}}}},\label{eq:tkA}
\end{equation}
where $\beta$ is a proportionality constant.

Before proceeding to present our numerical results, we note that according
to Eqs. \eqref{eq:JI} and \eqref{erttt}, it seems that a quantum
phase transition to ferromagnetic phase is possible in the effective
Hamailtonian of the system. However, it should be noticed also that
by choosing the appropriate parameters regime for deriving the effective
Hamiltonian in the above calculations, the system is certainly in
the Kondo regime and could not cross over to a ferromagnetic phase
within their corresponding parameters regime\cite{Yoo2014}.

\section{\label{sec:Numerical-renormalization-group}Numerical renormalization
group results}

Here, we provide numerical results obtained by numerical renormalization
group(NRG) method to confirm the manifestation of the Kondo effect
in the DQD-Qubit system. The NRG results are obtained using the ``\noun{NRG
Ljubljana}'' package\cite{zitko} by solving the total Hamiltonian
in Eq.\eqref{eq:tot_Ham} with $\Gamma=\pi\rho_{0}t^{2}=0.01D$, where
$\rho_{0}=1/\left(2D\right)$ is the density of states of the leads
which is assumed to be constant in the wide-band approximation and
$D=1$ is taken as the unit of the energy scales. Moreover, we take
$U=20\Gamma$ and $\Delta=10\Gamma$. We mention that for the chosen
values of $U$ and $\Delta$, the particular value of $\lambda$ at
which $U_{eff}$ will be vanished is $\lambda\approx28\Gamma$. Thus,
the system is expected to be in positive (negative) $U_{eff}$ regime
when $\lambda<28\Gamma$($\lambda>28\Gamma$).

\begin{figure}
\includegraphics[width=8.6cm]{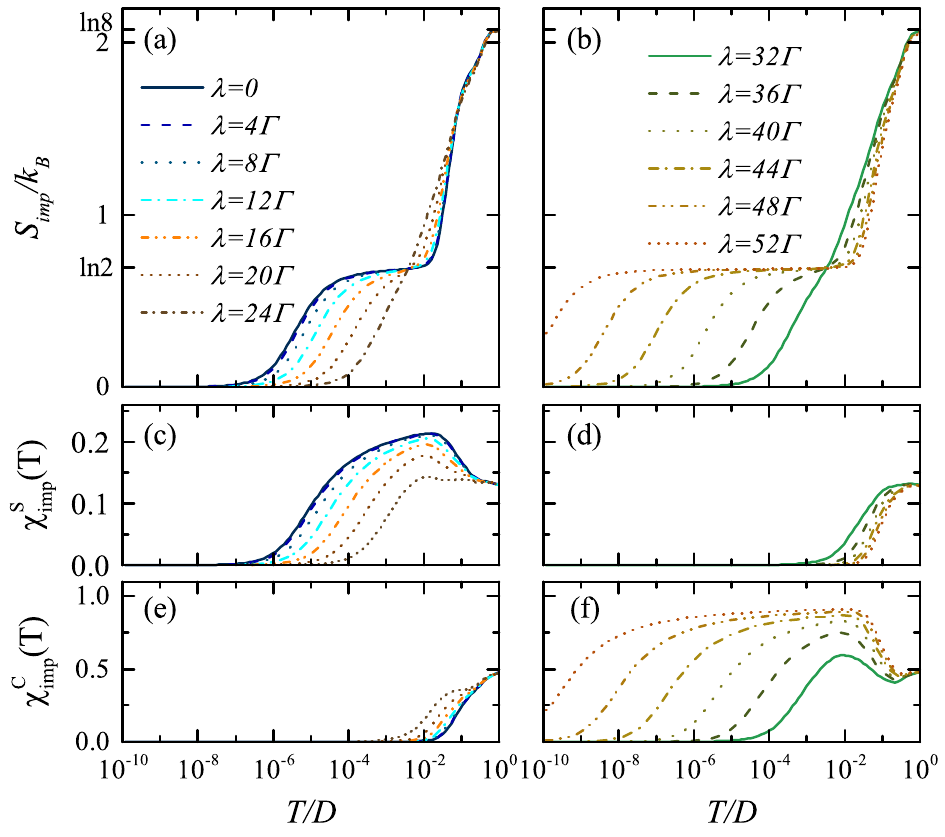}

\caption{\label{thermo} Impurity contribution to the total entropy(top panels),
magnetic susceptibility(middle panels) and charge susceptibility(bottom
panels) of the system. The values of $\lambda$ in the left and right
panels are corresponding to the positive and negative $U_{eff}$ regimes,
respectively. Other parameters are $\varepsilon_{1}=\varepsilon_{2}=-U/2,\,U=20\Gamma,\,\omega_{0}=\lambda\text{ and }\Delta=10\Gamma$.}
\end{figure}
First, we investigate the thermodynamic properties of the system.
In Fig.\ref{thermo}, we show the temperature dependence of the impurity
contribution to the entropy $S_{imp}(T)$, orbital pseudo-spin magnetic
susceptibility $\chi_{imp}^{S}(T)=(\bigl\langle(S^{z})^{2}\bigr\rangle-\bigl\langle(S^{z})^{2}\bigr\rangle_{0})/T$)
where $S^{z}=S_{d}^{z}+S_{c}^{z}$ and $\bigl\langle\ldots\bigr\rangle_{0}$
denotes the thermal expectation value in the absence of the DQD and
Qubit, and the total charge susceptibility $\chi_{imp}^{C}(T)=(\bigl\langle Q^{2}\bigr\rangle-\bigl\langle Q^{2}\bigr\rangle_{0})/T$
of the system in the p-h symmetric point and for several $\lambda$
values corresponding to the both $U_{eff}>0$ and $U_{eff}<0$ regimes.
As it is seen in Figs.\ref{thermo}(a) and (b), at high temperatures,
the system is in its free orbital fixed point where all $2^{3}$ states
of the DQD-Qubit system are equally probable and hence, the entropy
becomes $S_{imp}=k_{B}\ln8$. By decreasing the temperature, the system
crosses over to a local moment fixed point with $S_{imp}=k_{B}\ln2$
which means the presence of only two degrees of freedom in the system.
The nature of this local moment could be revealed by looking at the
respective magnetic and charge susceptibilities of the system which
are shown in Figs.\ref{thermo}(c)-(f). It is seen that, the local
moment in the $U_{eff}>0$ regime is accompanied by formation of an
orbital pseudo-spin magnetic moment in the system (see Fig.\ref{thermo}(c)).
On the other hand, in the $U_{eff}<0$ regime, we see that the established
local moment is of charge type (see Fig.\ref{thermo}(f)). Further
decrease of the temperature results in fully screening this local
moment after which the system crosses over to its Kondo strong coupling
fixed point with $S_{imp}=0$.

\begin{figure}
\includegraphics[width=8.6cm]{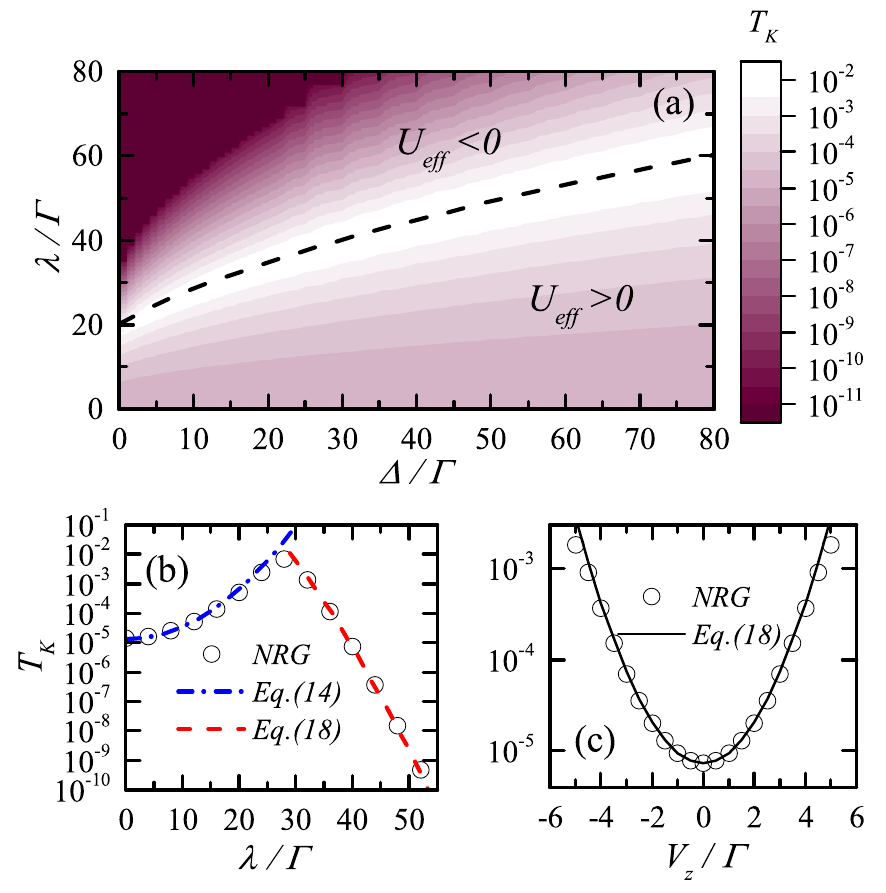}

\caption{\label{tk} (a) Kondo temperature of the system calculated from NRG
results with respect to the $\lambda$ and $\Delta$. Dashed line
shows the boundary between positive and negative $U_{eff}$ regimes.
(b) Comparison of Kondo temperature of the system for $\Delta=10\Gamma$,
calculated by NRG(circles), Eq.\eqref{eq:tkI}(dash-dotted line) and
Eq.\eqref{eq:tkA}(dashed line). (c) Kondo temperature of the system
as a function of $V_{z}$, calculated by NRG(circles), Eq.\eqref{eq:tkA}(dashed
line), for $\Delta=10\Gamma,\,\lambda=40\Gamma$. The proportionality
constants in Eqs.\eqref{eq:tkI} and \eqref{eq:tkA} are calculated
by numerical fitting to be $\alpha=0.034$ and $\beta=0.016$. Other
parameters are $\varepsilon_{1}=\varepsilon_{2}=-U/2,\,U=20\Gamma,\,\omega_{0}=\lambda$.}
\end{figure}
The particular temperature at which the local moment of the system
quenches in Figs.\ref{thermo}(a) and (b) is usually called the Kondo
temperature of the system. In Fig.\ref{tk}(a), we have shown $T_{K}$
values of the system calculated by the relation $S_{imp}\left(T_{K}\right)=0.5k_{B}\ln2$
from NRG results. The dashed line shows the points on which the values
of $U_{eff}$ vanish. It is seen that the behavior of $T_{K}$ values
in the positive $U_{eff}$ region differs from that in negative $U_{eff}$
region. This is more obvious in Fig.\ref{tk}(b), where we have plotted
the details of Fig.\ref{tk}(a) for a particular value of $\Delta$.
As we anticipated in Sec.\ref{sec:Low-energy-Effective-Hamiltonian},
in the positive $U_{eff}$ region, increasing $\lambda$ results in
an increase of the $T_{K}$ values, while in the negative $U_{eff}$
regime, the value of $T_{K}$ is drastically decreased by increasing
$\lambda$. In Fig.\ref{tk}(b), we have also compared the NRG results
with the results obtained using Eqs.\eqref{eq:tkI} and \eqref{eq:tkA},
where we can see very good agreement between the two results. To complete
our discussion about the Kondo temperature, in Fig.\ref{tk}(c), we
show the dependence of $T_{K}$ on the values of $V_{z}$ where $V_{z}=(\varepsilon_{1}-\varepsilon_{2})/2$,
and by assuming that the system is in the charge-Kondo regime. It
is apparent that departures from $V_{z}=0$ give rise to an enhancement
of $T_{K}$ which is again in agreement with the results obtained
using Eq.\eqref{eq:tkA}.

\begin{figure}
\includegraphics[width=8.6cm]{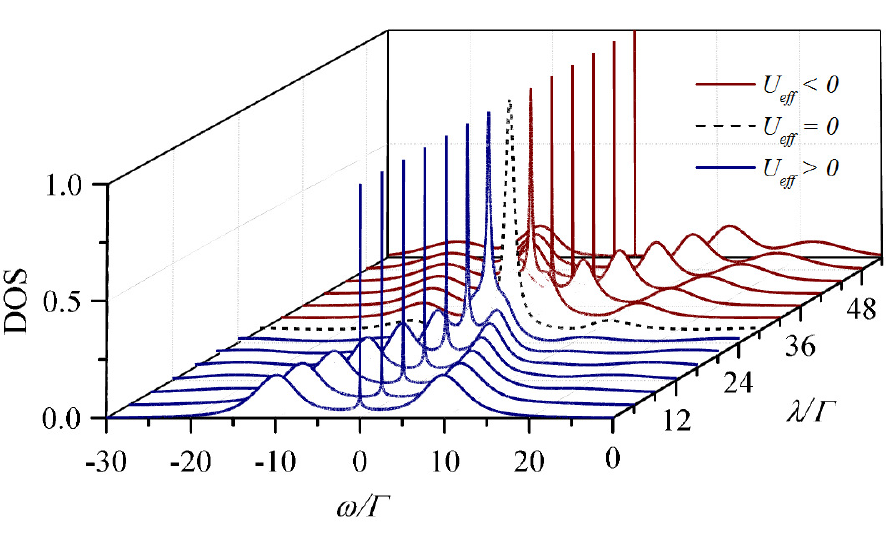}

\caption{\label{spectral}The spectral function(DOS) of the DQD for $\lambda$
values ranging from $\lambda=0$(orbital-Kondo regime) to $\lambda=52\Gamma$(charge-Kondo
regime). Dashed line shows the DOS for the particular $\lambda$ value
corresponding to the vanishing $U_{eff}$. Other parameters are $\varepsilon_{1}=\varepsilon_{2}=-U/2,\,U=20\Gamma,\,\omega_{0}=\lambda,\,\Delta=10\Gamma$.}
\end{figure}
Next, we consider spectral and transport properties of the DQD. In
Fig.\ref{spectral}, we show the spectral function of the DQD for
different values of $\lambda$. As it is obvious, when the system
is in orbital-Kondo regime, $\lambda<24\Gamma$, there is a central
Kondo peak along with two sideband peaks which are placed at the energies
corresponding to the single particle excitation energies of DQD. By
increasing $\lambda$, the system slightly crosses over to the charge-Kondo
regime which reflects in the central peak of the spectral function
by first transforming it to a lorentzian form around $\lambda\approx28\Gamma$
and then to a sharp charge-Kondo peak for $\lambda>32\Gamma$. Changing
$\lambda$ has also affected not only the energies but also the numbers
of the sideband peaks. This behavior can be explained by considering
the single particle excitation energies of the isolated DQD-Qubit
which could be calculated by using Eq.\eqref{eq:eigenenergy}. In
the orbital-Kondo regime, the ground state is the singly occupied
states. Thus, there are four possible sideband peaks in this regime
with energies equal to $\pm|E_{-}^{1,0}-E_{-}^{0,0}|$ and $\pm|E_{-}^{1,0}-E_{+}^{0,0}|$,
which are associated with the single particle excitation in DQD without
and with excitation of Qubit, respectively. However, as we can see
in Fig.\ref{spectral}, in orbital-Kondo regime, there are no sideband
peaks with Qubit excitation in the spectral function which is mainly
due to the weakness of the DQD-Qubit coupling. For larger $\lambda$
values, and specifically in the charge-Kondo regime, the sideband
peaks with Qubit excitation are much more visible. We note that in
the charge-Kondo regime, the ground state is composed of the empty
and doubly occupied states and therefore, the sideband peaks energies
are obtained from $\pm|E_{-}^{0,0}-E_{-}^{1,0}|$ and $\pm|E_{-}^{0,0}-E_{+}^{1,0}|$. 

\begin{figure}
\includegraphics[width=8.6cm]{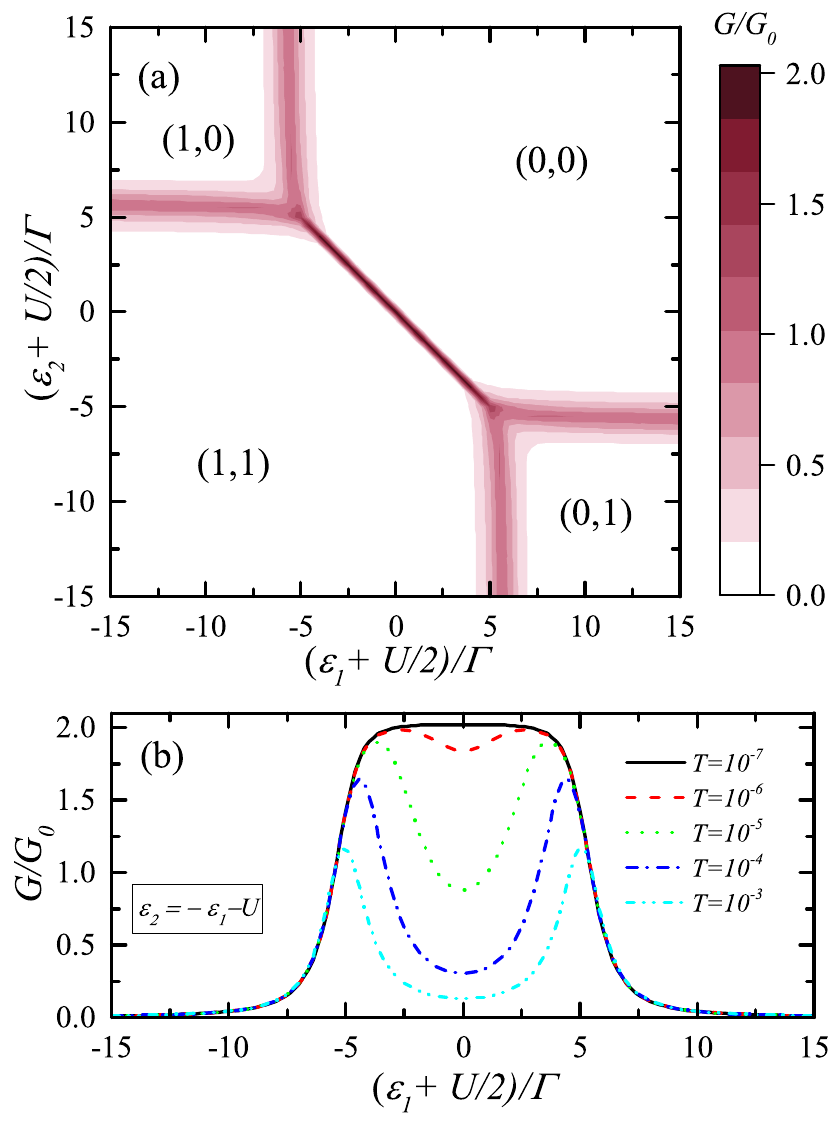}

\caption{\label{GE1E2} (a) Total linear conductance of the DQD with respect
to the $\varepsilon_{1}$ and $\varepsilon_{2}$ values. Charge configurations
of DQD are shown by ($n_{1}$,$n_{2}$) on the figure. (b) Temperature
dependence of the total linear conductance of DQD along the diagonal
line in (a). Other parameters are $U=20\Gamma,\,\Delta=10\Gamma,\,\omega_{0}=\lambda\text{ and }\lambda=40\Gamma$.}
\end{figure}
In Fig.\ref{GE1E2}(a), we show the total conductance of DQD $(G)$
in the $\varepsilon_{1}$-$\varepsilon_{2}$ plane. The parameters
are chosen such that the system is in the $U_{eff}<0$ regime. We
see that the profile of the total conductance is rotated with respect
to the usual charge stability diagrams expected for an interacting
parallel DQD\cite{van2002electron}. In particular, the presence of
a long degeneracy between the two charge configurations $(0,0)$ and
$(1,1)$ is a clue of an attractive interaction in the DQD. We can
also infer the strength of this attractive interaction as the length
of this degeneracy line segment between the two $(0,0)$ and $(1,1)$
regions. We have shown the temperature dependence of $G$ along this
segment in Fig.\ref{GE1E2}(b) in which we can see that the conductance
on this segment reaches the unitary value $G=2G_{0}=2e^{2}/h$ for
enough low temperatures while for higher temperatures, it is suppressed
except at the end points of the degeneracy segment. Hence, from these
observations, we can deduce that the nature of this unitary conductance
is of charge-Kondo type. It should be emphasized that a similar conductance
enhancement was reported in Ref.{[}\onlinecite{Hamo2016}{]}.

\section{\label{sec:Conclusions}Conclusions}

In conclusion, we considered a double-quantum-dot which is capacitively
coupled to a charge Qubit. It is shown that this capacitive coupling
renormalizes the inter-dot repulsive interaction in the double-quantum-dot
and in some situations makes it to be an attractive interaction between
the electrons in the double-quantum-dot. We found that appropriate
coupling of the double-quantum-dot with the electrodes could give
rise to two different types of the Kondo effect depending on the sign
of the inter-dot interaction in the double-quantum-dot. Namely, in
the positive interaction regime, the system shows an isotropic orbital-Kondo
effect while, in the negative interaction regime, an anisotropic charge-Kondo
effect is expected to be shown in the system. By deriving the low-energy
effective Hamiltonian of the system, we obtained the corresponding
Kondo exchange coupling constants as well as the characteristic Kondo
temperature of the system corresponding to these two different regimes.
Moreover, we employed the numerical renormalization group method to
confirm our analytical results and to extract some thermodynamic and
electronic transport properties of the system.
\begin{acknowledgments}
We would like to thank Farshad Ebrahimi for bringing the importance
of the model system to our attention. We are also grateful to Pablo
S. Cornaglia for useful comments and critical reading of a previous
version of the manuscript and Rok \v{Z}itko for helpful comments. 
\end{acknowledgments}

\appendix

\section{\label{sec:Derivation-of-effective}Derivation of effective Hamiltonian}

The low-energy effective Hamiltonian of the system can be calculated
using the method of degenerate perturbation theory. We will calculate
the second-order corrections to the unperturbed subsystem $\hat{\mathcal{H}}_{0}=\hat{\mathcal{H}}_{S}+\hat{\mathcal{H}}_{L}$
due to the perturbation coming from electron transitions between DQD
and electrodes which is described by $\hat{\mathcal{H}}_{T}$. First,
for future reference, we define eight projection operators to the
subspace of the eigenstates $\left|\psi_{\pm}^{n_{1},n_{2}}\right\rangle $
by \begin{subequations}

\begin{align}
\hat{P}_{\pm}^{0,0} & =\left(1-\hat{n}_{1}\right)\left(1-\hat{n}_{2}\right)\left|0,\pm\right\rangle \left\langle 0,\pm\right|,\\
\hat{P}_{\pm}^{1,0} & =\hat{n}_{1}\left(1-\hat{n}_{2}\right)\left|1,\pm\right\rangle \left\langle 1,\pm\right|,\\
\hat{P}_{\pm}^{0,1} & =\left(1-\hat{n}_{1}\right)\hat{n}_{2}\left|1,\pm\right\rangle \left\langle 1,\pm\right|,\\
\hat{P}_{\pm}^{1,1} & =\hat{n}_{1}\hat{n}_{2}\left|2,\pm\right\rangle \left\langle 2,\pm\right|.
\end{align}
\end{subequations}Now, the correction to the $\hat{\mathcal{H}}_{0}$
to the lowest order in $\hat{\mathcal{H}}_{T}$ can be calculated
by\cite{lim2010transport,garate2011charge}
\begin{equation}
\hat{\mathcal{H}}_{eff}=\sum_{\mathclap{\substack{n_{1},n_{2}=0,1\\
\nu=\pm
}
}}\frac{\hat{P}_{0}\hat{\mathcal{H}}_{T}\hat{P}_{\nu}^{n_{1},n_{2}}\hat{\mathcal{H}}_{T}\hat{P}_{0}}{E_{0}-E_{\nu}^{n_{1},n_{2}}},
\end{equation}
where $\hat{P}_{0}$ denotes the projection to the specific ground-state
of the unperturbed Hamiltonian $\hat{\mathcal{H}}_{0}$ with energy
$E_{0}$. Because we are interested in deriving the Kondo Hamiltonian
we restrict our calculations to the degenerate ground-states in the
following discussions.

\subsection{Positive $U_{eff}$ regime}

In this regime, the ground state becomes degenerate when $E_{-}^{1,0}=E_{-}^{0,1}$,
$E_{-}^{1,0}<E_{-}^{0,0}$ and $E_{-}^{1,0}<E_{-}^{1,1}$. Due to
the vast parameters in the system a large number of possible parameter
configurations can fulfill these conditions. By fixing $\lambda=\omega_{0}$,
which makes the Qubit to be particle-hole symmetric, we find that
the necessary conditions for degeneracy in the ground-state become
$\varepsilon_{1}=\varepsilon_{2}=V_{g}$ and $|\varepsilon_{1}+\frac{U}{2}|<\frac{1}{2}(U+\Omega_{1}-\Omega_{0})$,
where we have allowed for a gate voltage $V_{g}$ to consider the
general case.  By denoting the ground-state projector as $\hat{P}_{0}=\hat{P}_{-}^{1,0}+\hat{P}_{-}^{0,1}$,
we can evaluate the effective Hamiltonian in this case by\begin{widetext}

\begin{align}
\hat{\mathcal{H}}_{eff} & =\sum_{\mathclap{\substack{n_{1},n_{2}=0,1\\
\nu=\pm
}
}}\frac{\left\langle 1,-\right|\left(\hat{P}_{-}^{1,0}+\hat{P}_{-}^{0,1}\right)\hat{\mathcal{H}}_{T}\hat{P}_{\nu}^{n_{1},n_{2}}\hat{\mathcal{H}}_{T}\left(\hat{P}_{-}^{1,0}+\hat{P}_{-}^{0,1}\right)\left|1,-\right\rangle }{E_{-}^{1,0}-E_{\nu}^{n_{1},n_{2}}}.
\end{align}
Note that we have traced out the Qubit's degrees of freedom in order
to obtain a pure electronic effective Hamiltonian. Explicit calculation
of different terms in the above relation leads to\begin{subequations}

\begin{align}
\sum_{\nu=\pm}\frac{\left\langle 1,-\right|\left(\hat{P}_{-}^{1,0}+\hat{P}_{-}^{0,1}\right)\hat{\mathcal{H}}_{T}\hat{P}_{\nu}^{0,0}\hat{\mathcal{H}}_{T}\left(\hat{P}_{-}^{1,0}+\hat{P}_{-}^{0,1}\right)\left|1,-\right\rangle }{E_{-}^{1,0}-E_{\nu}^{0,0}} & =\frac{4t^{2}\left(-\Delta+V_{g}\right)}{4V_{g}\left(\Delta-V_{g}\right)+\lambda^{2}}\times\nonumber \\
\sum_{i=1,2}\Biggl[-\delta_{q,p}\left(\hat{n}_{i}-\hat{n}_{1}\hat{n}_{2}\right)+\hat{\tilde{c}}_{q,i}^{\dagger}\hat{\tilde{c}}_{p,i}\hat{n}_{i}+ & \hat{\tilde{c}}_{q,i}^{\dagger}\hat{\tilde{c}}_{p,\bar{i}}\hat{d}_{\bar{i}}^{\dagger}\hat{d}_{i}-\hat{\tilde{c}}_{q,i}^{\dagger}\hat{\tilde{c}}_{p,i}\hat{n}_{1}\hat{n}_{2}\Biggr],\\
\sum_{\nu=\pm}\frac{\left\langle 1,-\right|\left(\hat{P}_{-}^{1,0}+\hat{P}_{-}^{0,1}\right)\hat{\mathcal{H}}_{T}\hat{P}_{\nu}^{1,1}\hat{\mathcal{H}}_{T}\left(\hat{P}_{-}^{1,0}+\hat{P}_{-}^{0,1}\right)\left|1,-\right\rangle }{E_{-}^{1,0}-E_{\nu}^{1,1}} & =\frac{4t^{2}(\Delta+V_{g}+U)}{4(V_{g}+U)(\Delta+V_{g}+U)-\lambda^{2}}\times\nonumber \\
\sum_{i=1,2}\Biggl[-\hat{\tilde{c}}_{q,i}^{\dagger}\hat{\tilde{c}}_{p,i}\hat{n}_{\bar{i}} & +\hat{\tilde{c}}_{q,i}^{\dagger}\hat{\tilde{c}}_{p,\bar{i}}\hat{d}_{\bar{i}}^{\dagger}\hat{d}_{i}+\hat{\tilde{c}}_{q,i}^{\dagger}\hat{\tilde{c}}_{p,i}\hat{n}_{1}\hat{n}_{2}\Biggr].
\end{align}
\end{subequations}

Summing the above two contributions to the effective Hamiltonian and
using the identities $\hat{n}_{1}\hat{n}_{2}=1-\hat{n}_{d}$ and $\hat{n}_{d}=1$
which are valid only in the ground-state subspace of the positive
$U_{eff}$ regime, we can reach to the following expression for $\hat{\mathcal{H}}_{eff}$

\begin{align}
\hat{\mathcal{H}}_{eff} & =J\overrightarrow{S}_{d}.\overrightarrow{S}_{c}+K\sum_{\mathclap{\substack{i,j=1,2\\
q,p
}
}}\hat{\tilde{c}}_{q,i}^{\dagger}\hat{\tilde{c}}_{p,i}/2,
\end{align}
where $J$ and $K$ are given by\begin{subequations}
\begin{align}
J & =\frac{t^{2}}{\left(V_{g}+U\right)-\frac{\left(\lambda/2\right)^{2}}{\Delta+V_{g}+U}}-\frac{t^{2}}{V_{g}-\frac{\left(\lambda/2\right)^{2}}{V_{g}-\Delta}},\\
K & =\frac{t^{2}}{\left(V_{g}+U\right)-\frac{\left(\lambda/2\right)^{2}}{\Delta+V_{g}+U}}+\frac{t^{2}}{V_{g}-\frac{\left(\lambda/2\right)^{2}}{V_{g}-\Delta}}.
\end{align}
\end{subequations}Moreover, $\overrightarrow{S}_{d}=\frac{1}{2}\sum_{i,j=1,2}\hat{d}_{i}^{\dagger}\overrightarrow{\sigma}_{ij}\hat{d}_{j}$
is the corresponding pseudo-spin vector of DQD which is represented
by\begin{subequations}\label{Sdef}
\begin{gather}
S_{d}^{x}=\frac{1}{2}\left(\hat{d}_{2}^{\dagger}\hat{d}_{1}+\hat{d}_{2}^{\dagger}\hat{d}_{1}\right),\\
S_{d}^{y}=\frac{i}{2}\left(\hat{d}_{2}^{\dagger}\hat{d}_{1}-\hat{d}_{2}^{\dagger}\hat{d}_{1}\right),\\
S_{d}^{z}=\frac{1}{2}\left(\hat{n}_{1}-\hat{n}_{2}\right),
\end{gather}
\end{subequations} and, analogously, $\overrightarrow{S}_{c}=\frac{1}{2}\sum_{i,j=1,2}\sum_{p,q}\hat{\tilde{c}}_{q,i}^{\dagger}\overrightarrow{\sigma}_{ij}\hat{\tilde{c}}_{p,j}$
is the pseudo-spin vector for the electrodes.

\subsection{Negative $U_{eff}$ regime}

In this regime, a similar calculation as that of positive $U_{eff}$
regime but for the the ground state composed of the two states $\bigl|\psi_{-}^{0,0}\bigr\rangle$
and $\bigl|\psi_{-}^{1,1}\bigr\rangle$ leads to the conditions $\lambda=\omega_{0}$,
$\varepsilon_{1}+\varepsilon_{2}+U=0$ and $|V_{z}|<\frac{1}{2}(-U-\Omega_{1}+\Omega_{0})$
, where $V_{z}=\varepsilon_{1}+U/2=-(\varepsilon_{2}+U/2)$. Then,
the ground-state projector is $\hat{P}_{0}=\hat{P}_{-}^{0,0}+\hat{P}_{-}^{1,1}$,
and the effective Hamiltonian can be calculated by

\begin{align}
\hat{\mathcal{H}}_{eff} & =\sum_{\mathclap{\substack{n_{1},n_{2}=0,1\\
\nu=\pm
}
}}\frac{\left(\left\langle 0,-\right|\hat{P}_{-}^{0,0}+\left\langle 2,-\right|\hat{P}_{-}^{1,1}\right)\hat{\mathcal{H}}_{T}\hat{P}_{\nu}^{n_{1},n_{2}}\hat{\mathcal{H}}_{T}\left(\hat{P}_{-}^{0,0}\left|0,-\right\rangle +\hat{P}_{-}^{1,1}\left|2,-\right\rangle \right)}{E_{-}^{0,0}-E_{\nu}^{n_{1},n_{2}}}.
\end{align}
By calculating the summations in the above relation, after some algebra,
we get\begin{subequations}

\begin{align}
\sum_{\nu=\pm}\frac{\left\langle 0,-\right|\hat{P}_{-}^{0,0}\hat{\mathcal{H}}_{T}\hat{P}_{\nu}^{1,0}\hat{\mathcal{H}}_{T}\hat{P}_{-}^{0,0}\left|0,-\right\rangle }{E_{-}^{0,0}-E_{\nu}^{1,0}}+ & \frac{\left\langle 2,-\right|\hat{P}_{-}^{1,1}\hat{\mathcal{H}}_{T}\hat{P}_{\nu}^{1,0}\hat{\mathcal{H}}_{T}\hat{P}_{-}^{1,1}\left|2,-\right\rangle }{E_{-}^{0,0}-E_{\nu}^{1,0}}=\nonumber \\
\frac{2t^{2}}{\Delta^{2}-\left(-U+2V_{z}+\Omega_{0}\right){}^{2}} & \left(\lambda-U+2V_{z}+\frac{\Delta^{2}\left(\lambda+2\Omega_{0}\right)}{\left(\Delta^{2}+\lambda^{2}+\lambda\Omega_{0}\right)}\right)\times\nonumber \\
\Biggl[\hat{\tilde{c}}_{q,1}^{\dagger}\hat{\tilde{c}}_{p,1}-\hat{\tilde{c}}_{q,1}^{\dagger}\hat{\tilde{c}}_{p,1}\hat{n}_{d}+ & \delta_{q,p}\hat{n}_{1}\hat{n}_{2}+\left(\hat{\tilde{c}}_{q,1}^{\dagger}\hat{\tilde{c}}_{p,1}-\hat{\tilde{c}}_{q,2}^{\dagger}\hat{\tilde{c}}_{p,2}\right)\hat{n}_{1}\hat{n}_{2}\Biggr],\\
\sum_{\nu=\pm}\frac{\left\langle 0,-\right|\hat{P}_{-}^{0,0}\hat{\mathcal{H}}_{T}\hat{P}_{\nu}^{0,1}\hat{\mathcal{H}}_{T}\hat{P}_{-}^{0,0}\left|0,-\right\rangle }{E_{-}^{0,0}-E_{\nu}^{0,1}}+ & \frac{\left\langle 2,-\right|\hat{P}_{-}^{1,1}\hat{\mathcal{H}}_{T}\hat{P}_{\nu}^{0,1}\hat{\mathcal{H}}_{T}\hat{P}_{-}^{1,1}\left|2,-\right\rangle }{E_{-}^{0,0}-E_{\nu}^{0,1}}=\nonumber \\
\frac{2t^{2}}{\Delta^{2}-\left(U+2V_{z}-\Omega_{0}\right){}^{2}} & \left(\lambda-U-2V_{z}+\frac{\Delta^{2}\left(\lambda+2\Omega_{0}\right)}{\left(\Delta^{2}+\lambda^{2}+\lambda\Omega_{0}\right)}\right)\times\nonumber \\
\Biggl[\hat{\tilde{c}}_{q,2}^{\dagger}\hat{\tilde{c}}_{p,2}-\hat{\tilde{c}}_{q,2}^{\dagger}\hat{\tilde{c}}_{p,2}\hat{n}_{d}+ & \delta_{q,p}\hat{n}_{1}\hat{n}_{2}+\left(\hat{\tilde{c}}_{q,2}^{\dagger}\hat{\tilde{c}}_{p,2}-\hat{\tilde{c}}_{q,1}^{\dagger}\hat{\tilde{c}}_{p,1}\right)\hat{n}_{1}\hat{n}_{2}\Biggr],\\
\sum_{\nu=\pm}\frac{\left\langle 0,-\right|\hat{P}_{-}^{0,0}\hat{\mathcal{H}}_{T}\hat{P}_{\nu}^{0,1}\hat{\mathcal{H}}_{T}\hat{P}_{-}^{1,1}\left|2,-\right\rangle }{E_{-}^{0,0}-E_{\nu}^{0,1}}+ & \frac{\left\langle 2,-\right|\hat{P}_{-}^{1,1}\hat{\mathcal{H}}_{T}\hat{P}_{\nu}^{0,1}\hat{\mathcal{H}}_{T}\hat{P}_{-}^{0,0}\left|0,-\right\rangle }{E_{-}^{0,0}-E_{\nu}^{0,1}}=\nonumber \\
 & \frac{2\Delta t^{2}/\Omega_{0}}{2V_{z}-U+\frac{\lambda^{2}}{2V_{z}-U+2\Omega_{0}}}\sum_{i=1,2}\Biggl[\hat{\tilde{c}}_{q,i}^{\dagger}\hat{\tilde{c}}_{p,\bar{i}}^{\dagger}\hat{d}_{\bar{i}}\hat{d}_{i}+h.c.\Biggr],\\
\sum_{\nu=\pm}\frac{\left\langle 0,-\right|\hat{P}_{-}^{0,0}\hat{\mathcal{H}}_{T}\hat{P}_{\nu}^{1,0}\hat{\mathcal{H}}_{T}\hat{P}_{-}^{1,1}\left|2,-\right\rangle }{E_{-}^{0,0}-E_{\nu}^{1,0}}+ & \frac{\left\langle 2,-\right|\hat{P}_{-}^{1,1}\hat{\mathcal{H}}_{T}\hat{P}_{\nu}^{1,0}\hat{\mathcal{H}}_{T}\hat{P}_{-}^{0,0}\left|0,-\right\rangle }{E_{-}^{0,0}-E_{\nu}^{1,0}}=\nonumber \\
 & \frac{-2\Delta t^{2}/\Omega_{0}}{2V_{z}+U+\frac{\lambda^{2}}{2V_{z}+U-2\Omega_{0}}}\sum_{i=1,2}\Biggl[\hat{\tilde{c}}_{q,i}^{\dagger}\hat{\tilde{c}}_{p,\bar{i}}^{\dagger}\hat{d}_{\bar{i}}\hat{d}_{i}+h.c.\Biggr].
\end{align}
\end{subequations}

Now, by defining the iso-spin operators $\overrightarrow{I}_{d}$
of the DQD as\begin{subequations}\label{Idef}
\begin{gather}
I_{d}^{x}=\frac{1}{2}\left(\hat{d}_{2}^{\dagger}\hat{d}_{1}^{\dagger}+\hat{d}_{1}\hat{d}_{2}\right),\\
I_{d}^{y}=\frac{i}{2}\left(\hat{d}_{2}^{\dagger}\hat{d}_{1}^{\dagger}-\hat{d}_{1}\hat{d}_{2}\right),\\
I_{d}^{z}=\frac{1}{2}\left(\hat{n}_{d}-1\right),
\end{gather}
\end{subequations}with a similar definition for the $\overrightarrow{I}_{c}$
as the iso-spin operators of the electrodes and using the identities
$2\hat{n}_{1}\hat{n}_{2}=\hat{n}_{d}$ and $\hat{n}_{1}=\hat{n}_{2}$
which are valid in the ground state of the negative $U_{eff}$ regime,
we can obtain the effective Hamiltonian in Eq.\eqref{eq:heffA}.

\end{widetext}

\bibliographystyle{apsrev}
\bibliography{extraref}

\end{document}